\documentclass[prd,nofootinbib,reprint,twocolumn,showpacs,superscriptaddress,floatfix]{revtex4-1}
\pdfoutput=1 
\usepackage{bm, color} 
\usepackage{amssymb,amsfonts,slashed,amsthm,amsmath, graphicx, soul, empheq, cancel}
\usepackage{hyperref}
\usepackage[caption=false]{subfig}

\begin{document}

\newcommand{\vev}[1]{ \left\langle {#1} \right\rangle }
\newcommand{\bra}[1]{ \langle {#1} | }
\newcommand{\ket}[1]{ | {#1} \rangle }
\newcommand{\eV}{ \ {\rm eV} }
\newcommand{\KeV}{ \ {\rm keV} }
\newcommand{\MeV}{\  {\rm MeV} }
\newcommand{\GeV}{\  {\rm GeV} }
\newcommand{\TeV}{\  {\rm TeV} }
\newcommand{\1}{\mbox{1}\hspace{-0.25em}\mbox{l}}
\newcommand{\Red}[1]{{\color{red} {#1}}}

\newcommand{\lmk}{\left(}  
\newcommand{\rmk}{\right)}
\newcommand{\lkk}{\left[}  
\newcommand{\rkk}{\right]}
\newcommand{\lhk}{\left \{ }  
\newcommand{\rhk}{\right \} }
\newcommand{\del}{\partial}  
\newcommand{\la}{\left\langle} 
\newcommand{\ra}{\right\rangle}
\newcommand{\half}{\frac{1}{2}}

\newcommand{\bea}{\begin{array}}
\newcommand{\eea}{\end{array}}
\newcommand{\beq}{\begin{eqnarray}}
\newcommand{\eeq}{\end{eqnarray}}
\newcommand{\eq}[1]{Eq.~(\ref{#1})}

\newcommand{\dd}{\mathrm{d}}
\newcommand{\Mpl}{M_{\rm Pl}}
\newcommand{\mg}{m_{3/2}}
\newcommand{\abs}[1]{\left\vert {#1} \right\vert}
\newcommand{\mphi}{m_{\phi}}
\newcommand{\Hz}{\ {\rm Hz}}
\newcommand{\for}{\quad \text{for }}
\newcommand{\Min}{\text{Min}}
\newcommand{\Max}{\text{Max}}
\newcommand{\Kahler}{K\"{a}hler }
\newcommand{\cphi}{\varphi}
\newcommand{\Tr}{\text{Tr}}
\newcommand{\diag}{{\rm diag}}

\newcommand{\SUf}{SU(3)_{\rm f}}
\newcommand{\Upq}{U(1)_{\rm PQ}}
\newcommand{\Zpq}{Z^{\rm PQ}_3}
\newcommand{\Cpq}{C_{\rm PQ}}
\newcommand{\ubar}{u^c}
\newcommand{\dbar}{d^c}
\newcommand{\ebar}{e^c}
\newcommand{\nubar}{\nu^c}
\newcommand{\Ndw}{N_{\rm DW}}
\newcommand{\Fpq}{F_{\rm PQ}}
\newcommand{\fpq}{v_{\rm PQ}}
\newcommand{\Br}{{\rm Br}}
\newcommand{\Lag}{\mathcal{L}}
\newcommand{\Lqcd}{\Lambda_{\rm QCD}}

\newcommand{\ji}{j_{\rm inf}} 
\newcommand{\jb}{j_{B-L}} 
\newcommand{\M}{M} 
\newcommand{\im}{{\rm Im} }
\newcommand{\re}{{\rm Re} }

\def\lrf#1#2{ \left(\frac{#1}{#2}\right)}
\def\lrfp#1#2#3{ \left(\frac{#1}{#2} \right)^{#3}}
\def\lrp#1#2{\left( #1 \right)^{#2}}
\def\REF#1{Ref.~\cite{#1}}
\def\SEC#1{Sec.~\ref{#1}}
\def\FIG#1{Fig.~\ref{#1}}
\def\EQ#1{Eq.~(\ref{#1})}
\def\EQS#1{Eqs.~(\ref{#1})}
\def\TEV#1{10^{#1}{\rm\,TeV}}
\def\GEV#1{10^{#1}{\rm\,GeV}}
\def\MEV#1{10^{#1}{\rm\,MeV}}
\def\KEV#1{10^{#1}{\rm\,keV}}
\def\blue#1{\textcolor{blue}{#1}}
\def\red#1{\textcolor{blue}{#1}}

\newcommand{\eff}{\Delta N_{\rm eff}}
\newcommand{\neff}{\Delta N_{\rm eff}}
\newcommand{\cc}{\Omega_\Lambda}
\newcommand{\Mpc}{\ {\rm Mpc}}
\newcommand{\Msolar}{M_\odot}

\def\sn#1{\textcolor{red}{#1}}
\def\SN#1{\textcolor{red}{[{\bf SN:} #1]}}
\def\my#1{\textcolor{blue}{#1}}
\def\MY#1{\textcolor{blue}{[{\bf MY:} #1]}}

\begin{flushright}
TU-1261
\end{flushright}

\title{
Interpreting Cosmic Birefringence and DESI Data with Evolving Axion in $\Lambda$CDM
}

\author{Shota Nakagawa}
\affiliation{Tsung-Dao Lee Institute, Shanghai Jiao Tong University, \\
No.~1 Lisuo Road, Pudong New Area, Shanghai 201210, China}
\affiliation{School of Physics and Astronomy, Shanghai Jiao Tong University, \\
800 Dongchuan Road, Shanghai 200240, China}

\author{Yuichiro Nakai}
\affiliation{Tsung-Dao Lee Institute, Shanghai Jiao Tong University, \\
No.~1 Lisuo Road, Pudong New Area, Shanghai 201210, China}
\affiliation{School of Physics and Astronomy, Shanghai Jiao Tong University, \\
800 Dongchuan Road, Shanghai 200240, China}

\author{Yu-Cheng Qiu}
\affiliation{Tsung-Dao Lee Institute, Shanghai Jiao Tong University, \\
No.~1 Lisuo Road, Pudong New Area, Shanghai 201210, China}
\affiliation{School of Physics and Astronomy, Shanghai Jiao Tong University, \\
800 Dongchuan Road, Shanghai 200240, China}

\author{Masaki Yamada}
\affiliation{Department of Physics, Tohoku University, \\
6-3 Azaaoba, Aramaki, Aoba-ku, Sendai 980-8578, Japan}

\begin{abstract}

Recent cosmological observations have revealed growing tensions with the standard $\Lambda$CDM model, including indications of isotropic cosmic birefringence and deviations from $w = -1$ in the dark energy equation of state, as suggested by DESI and supernova measurements. In this paper, we point out that such deviations can arise even from a subdominant energy density component. We then propose a unified framework based on a dynamical axion field that simultaneously accounts for both anomalies, providing a simple and natural extension of the standard $\Lambda$CDM model. In our scenario, the axion field with $2H_0\lesssim m\lesssim 7H_0$, where $H_0$ is the current Hubble constant, induces a nonzero rotation of the CMB polarization plane and modifies the present-day dark energy equation of state. This framework accommodates recent observational data with natural parameter choices, even for a string axion with a decay constant of order $10^{17}\,$GeV.

\end{abstract}

\maketitle


\section{Introduction
\label{sec:introduction}}

The standard cosmological model, \(\Lambda\)CDM, has successfully described the large-scale structure and evolution of the Universe. However, growing observational evidence suggests potential deviations from this paradigm.
Two such possible deviations, \emph{isotropic cosmic birefringence} and \emph{evolving dark energy}, are currently under active investigation in light of recent data.

Cosmic birefringence~\cite{Carroll:1989vb,Carroll:1991zs,Harari:1992ea}
refers to the in-flight rotation of the polarization plane of cosmic microwave background (CMB) photons as they propagate through the Universe.
The overall rotation angle from the last scattering surface to the present is called the isotropic cosmic birefringence (ICB) angle,
denoted as $\beta$.
The analysis of CMB polarization data has revealed the indication of a nonzero rotation angle
\cite{Minami:2020odp,Diego-Palazuelos:2022dsq,Eskilt:2022wav,Eskilt:2022wav,Eskilt:2022cff},
\begin{align}
\label{obsCB}
    \beta= 0.34^{\circ} \pm 0.09^{\circ} .
\end{align}
Moreover, the recent observational result of the Atacama Cosmology Telescope has provided additional support~\cite{ACT:2025fju}:
\begin{align}
\label{obsCB2}
    \beta= 0.20^{\circ} \pm 0.08^{\circ} ,
\end{align}
strengthening the case that the ICB may be a real physical effect.
Since the Standard Model and its extensions involving higher-dimensional operators composed solely of Standard Model fields cannot account for this phenomenon
\cite{Nakai:2023zdr},
the observed ICB may indicate a parity-violating interaction of a new light field beyond the Standard Model. 
A particularly compelling candidate for such a field is a background axion-like field coupled to photons
\cite{Pospelov:2008gg,Finelli:2008jv,Panda:2010uq,Lee:2013mqa,Zhao:2014yna,Liu:2016dcg,Sigl:2018fba,Fedderke:2019ajk,Fujita:2020ecn,Takahashi:2020tqv,Fung:2021wbz,Nakagawa:2021nme,Jain:2021shf,Choi:2021aze,Obata:2021nql,Nakatsuka:2022epj,Lin:2022niw,Gasparotto:2022uqo,Lee:2022udm,Jain:2022jrp,Murai:2022zur,Gonzalez:2022mcx,Qiu:2023los,Eskilt:2023nxm,Namikawa:2023zux,Gasparotto:2023psh,Luo:2023cxo,Ferreira:2023jbu,Greco:2024oie,Tada:2024znt,Naokawa:2024xhn,Murai:2024yul,Zhang:2024dmi}.

Recent observations suggest that the dark energy (DE) may not be a cosmological constant. 
Baryon Acoustic Oscillation data from the Dark Energy Spectroscopic Instrument (DESI)~\cite{DESI:2024mwx}, combined with type Ia supernova measurements from Union3~\cite{Rubin:2023ovl} and Year-5 data from the Dark Energy Survey (DES)~\cite{DES:2024jxu}, point toward a time-varying DE equation of state. 
The data prefer a scenario in which \( w < -1 \) at earlier times and \( w > -1 \) at the present epoch.
This behavior is further supported by the latest DESI Data Release 2 (DR2) results~\cite{DESI:2025zgx,Lodha:2025qbg}. 
However, it has been discussed that a ``phantom" phase with \( w < -1 \) may be an artifact of prior assumptions
\cite{Cortes:2024lgw}, 
which motivates us to focus on the possibility that $w > -1$ at present. 

An axion-like field has been proposed as a candidate for time-varying DE.
In particular, a quintessential axion has been studied in the literature
\cite{Fukugita:1994hq,Frieman:1995pm,Kim:1998kx,Kim:1999dc,Choi:1999xn,Nomura:2000yk,Kim:2002tq,Copeland:2006wr,Panda:2010uq,Choi:2021aze,Obata:2021nql,Gasparotto:2022uqo,Qiu:2023los,Tada:2024znt,Berbig:2024aee}
as a unified explanation for both ICB and evolving DE, without invoking a cosmological constant.\footnote{
Non-axion scalar fields can also serve as evolving DE candidates to account for the DESI result (see Refs.~\cite{Copeland:2006wr,Linder:2007wa,Linder:2010py,Chakraborty:2025syu} and references therein).}
However, as emphasized in Ref.~\cite{Choi:2021aze}, such models typically require a relatively large axion-photon coupling.
This is because the axion field displacement from the recombination epoch to today is limited by a small axion mass required for DE evolution.
As a result, achieving the observed ICB in these models necessitates enhancing the axion-photon coupling beyond a natural {$\mathcal O (1)$} value.
Moreover, the observed DE density necessitates an axion decay constant near the Planck scale, in tension with the expected values for string axions.

In this paper, we propose an extension of the $\Lambda$CDM model that addresses both ICB and evolving DE in a unified and natural framework.
We introduce a dynamical axion-like field while retaining a nonzero cosmological constant, $a\Lambda$CDM for short. 
The axion mass is assumed to be slightly larger than the present Hubble parameter, such that the axion begins to evolve just prior to the present epoch.
This late-time dynamics leads to a sizable birefringence angle and a deviation in the dark energy equation of state from $-1$. 
Although the axion energy density remains subdominant compared to the cosmological constant, we show that its impact on the expansion rate is sufficient to account for
the recent conclusion that DE equation of state $w>-1$ today.
Unlike models without a vacuum energy, our framework allows for a smaller axion decay constant, well below the Planck scale, which is favorable for a string axion.%
\footnote{
However, recent studies suggest that the axion-photon coupling invoked to explain the ICB may not be consistent with heterotic string theories or simple 4D GUTs~\cite{Agrawal:2022lsp,Agrawal:2024ejr}. 
}
Our model is economical in the sense that a single axion-like field accounts for both anomalies, potentially signaling deviations from the standard $\Lambda$CDM cosmology.

The structure of the paper is as follows.
In Section~\ref{matter_domination}, we present the theoretical framework of our model.
Section~\ref{mass_model} explores the parameter space compatible with current observations.
Section~\ref{sec:Discussion} is devoted to conclusions and discussion.

\section{$\Lambda$CDM with an axion}
\label{matter_domination}

We consider an extension of the $\Lambda$CDM model by introducing an ultralight axion field.
The axion potential is taken to be the standard sine-Gordon form such as 
\beq
 V(\phi) 
 &=& m^2 f_\phi^2 \lmk 1 -  \cos \frac{\phi}{f_\phi} \rmk  .
\eeq
The inflation dilutes the gradient energy density for the axion field, $\phi(\vec x) = \phi_{\rm ini}$. So we only have to consider its time evolution.
The equation of motion for $\phi$ is given by 
\beq
 \ddot{\phi} + 3 H(t) \dot{\phi}  + V'(\phi)= 0 \ , 
 \label{eom}
\eeq
where $t$ is the cosmic time and the Hubble parameter is 
\beq
H^2=H^2_0\left(\Omega_{\rm{rad}}a^{-4}+\Omega_{\rm{mat}}a^{-3}+\Omega_\Lambda + \frac{\rho_{\phi}(a)}{\rho_c}\right) \ .
\label{eq:H}
\eeq 
$H_0\sim 10^{-33}$~eV is the Hubble constant today.
Here, we include a constant vacuum energy, or cosmological constant, denoted by $\Lambda$. 
The fractional density parameters are defined as $\Omega_i\equiv \rho_i/\rho_c$ for $i={\rm rad, \, mat}, \, \Lambda$, where $\rho_c$ is the critical energy density today. 
The energy density of the axion is given by $\rho_\phi = \dot{\phi}^2/2 + V(\phi)$.
We treat the initial misalignment angle $\theta_{\rm ini} \equiv \phi_{\rm ini}/f_\phi$ as an $\mathcal{O}(1)$ free parameter.

In our scenario, the ultralight axion is still in the initial phase of its evolution toward the minimum of its potential at present. Consequently, the kinetic energy is subdominant relative to the potential energy, enabling the field to contribute effectively to dark energy.
The cosmological constant remains the dominant component of the present-day energy density, while the axion field contributes as a subdominant component.
However, due to its nontrivial dynamics, the axion can still affect the expansion history of the Universe, and it modifies the DE equation of state.

The effective equation of state for DE, in the presence of vacuum energy and the axion field, is given by 
\beq
 w = \frac{\dot{\phi}^2/2-V(\phi) - \Lambda}{\dot{\phi}^2/2+V(\phi) + \Lambda}  \ .
\eeq
We assume that the axion mass $m$ is of the order of the present-day Hubble parameter $H_0$, such that the axion dynamics becomes relevant at late times. 
The axion decay constant 
$f_\phi$ is taken to be of order the string scale $\sim 10^{17} \, \rm GeV$, as motivated by the string axion.
Under these assumptions, the axion potential energy $V(\phi)$ remains subdominant but can still be comparable to the current total energy density, thus modifying the expansion rate of the Universe.

We further assume that the axion field $\phi$ couples to the Standard Model photon through the interaction,
\beq
 {\cal L} 
 \supset -c_\gamma \frac{\alpha}{4 \pi} \frac{\phi}{f_\phi} F_{\mu \nu} \tilde{F}^{\mu \nu} 
 \equiv -\frac{1}{4} g_{\phi \gamma \gamma} \phi F_{\mu \nu} \tilde{F}^{\mu \nu} \ , 
\eeq
where $\alpha$ is the fine-structure constant, $c_\gamma$ is the $U(1)_{\rm EM}$ anomaly coefficient, and $g_{\phi\gamma\gamma}$ denotes the axion-photon coupling constant. 
Here, $F_{\mu\nu}$ and $\tilde{F}_{\mu\nu}$ represent the electromagnetic field strength tensor and its dual, respectively.

If the axion evolves after the recombination epoch, it induces ICB, leading to a rotation of the CMB polarization plane. The rotation angle is given by~\cite{Harari:1992ea}
\beq
 \beta \simeq 0.42^{\circ}\, \times 
 \lmk
 c_\gamma \frac{\Delta \phi}{2 \pi f_\phi} \rmk.
 \label{obs}
\eeq
where $\Delta \phi \equiv \phi_{\rm p} - \la \phi_{\rm LSS} \ra$ with $\la \phi_{\rm LSS} \ra$ denoting the average value of the axion field at the last scattering surface, and $\phi_{\rm p}$ the present field value. 
In our case, $\la \phi_{\rm LSS} \ra \simeq \phi_{\rm ini}$. 

The evolution of the axion field can be solved by coupled equations Eq.~\eqref{eom} and Eq.~\eqref{eq:H}, which are determined by parameters $\{\Omega_\Lambda, m, f_\phi,\theta_{\rm ini}\}$. To have a consistent cosmic history, one has to ensure that axion and $\Lambda$ account for observed DE density today, which means that $\Omega_\Lambda + \rho_\phi/\rho_c \simeq 0.7$ today. Once the axion field evolution is solved, together with $c_\gamma$, the ICB can be explained.

\begin{figure}[t]  
    \includegraphics[width=1.\linewidth]{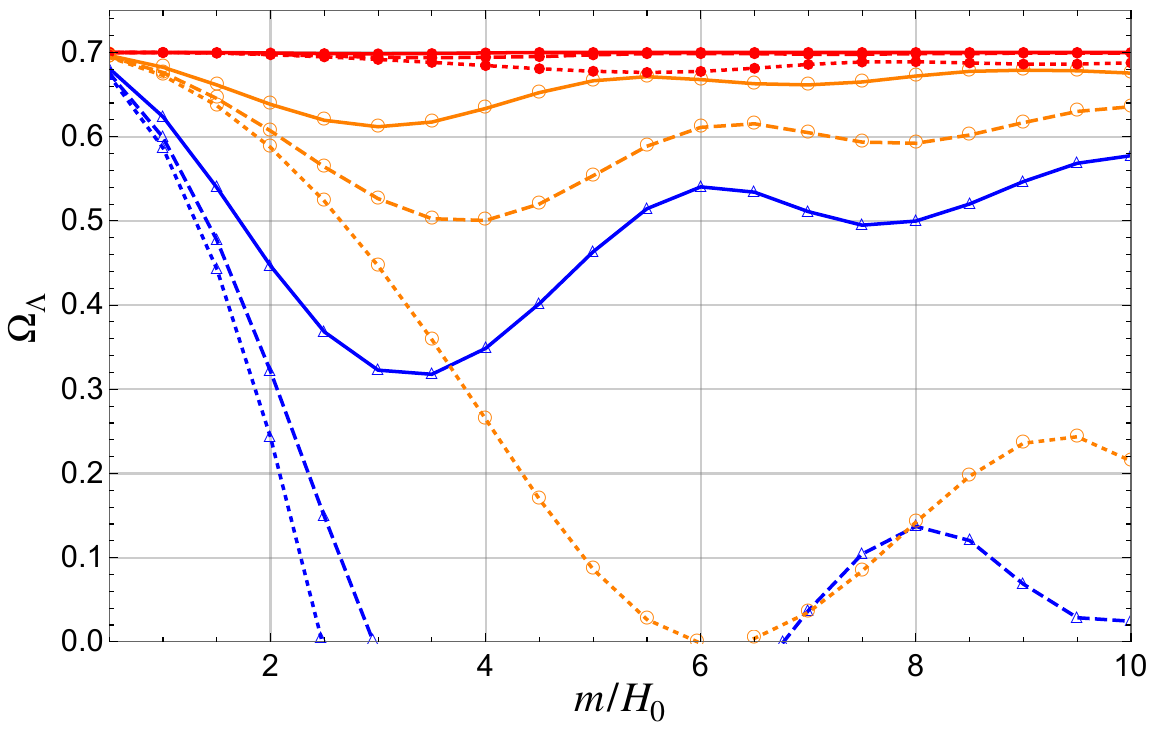}
	\caption{
    The abundance of the cosmological constant as a function of mass. 
    The red ($\bullet$), orange ($\circ$), and blue ($\triangle$) lines correspond to $f_\phi[10^{17}\GeV] = 1,~5,~10$, respectively.  
    The solid, dashed, and dotted lines indicate $\theta_{\rm ini} = 2,~2.5,~3$, respectively.
    }
	\label{fig:DEratio}
\end{figure}

Let us briefly summarize the abundance ratio of DE as shown in \FIG{fig:DEratio}.
In this figure, we plot the abundance of the cosmological constant, assuming $\Omega_{\Lambda} +\rho_\phi(a_0)/\rho_c \simeq 0.7$
for various values of $f_\phi$ and $\theta_{\rm ini}$, with $a_0$ being the present scale factor.
The red, orange, and blue lines correspond to 
$f_\phi[10^{17}\GeV]=1,~5,~10$, respectively. The solid, dashed, and dotted lines corresponds to 
$\theta_{\rm ini}=2,~2.5,~3$, respectively.
In the small mass regime, $m \sim H_0$, the axion energy remains subdominant compared to the cosmological constant.
As the axion mass increases, its energy density grows, particularly for larger values of $f_\phi$ and $\theta_{\rm ini}$. However, even in this case, the kinetic energy remains subdominant, as mentioned above.
Our region of interest lies around the global minima in \FIG{fig:DEratio} for each $(f_\phi, \theta_{\rm ini})$ combination. In this region, dynamical DE is realized, as will be shown in the lower panel of \FIG{fig:ICB}.
In the regime of even larger axion mass, the axion begins to oscillate slowly at present,%
\footnote{This is evident from the oscillatory behavior seen in the high-mass region of \FIG{fig:DEratio}.}
leading to a suppression of the axion energy. 
As we will see shortly, the ICB can be explained when $m$ is around or exceeds the mass value at the global minimum.
Finally, we note that there are cases with $\Omega_\Lambda = 0$, which align with the quintessential axion scenario where the cosmological constant is absent.

\section{Results}
\label{mass_model}

\begin{figure}[t]
	\centering	\includegraphics[width=1.\linewidth]{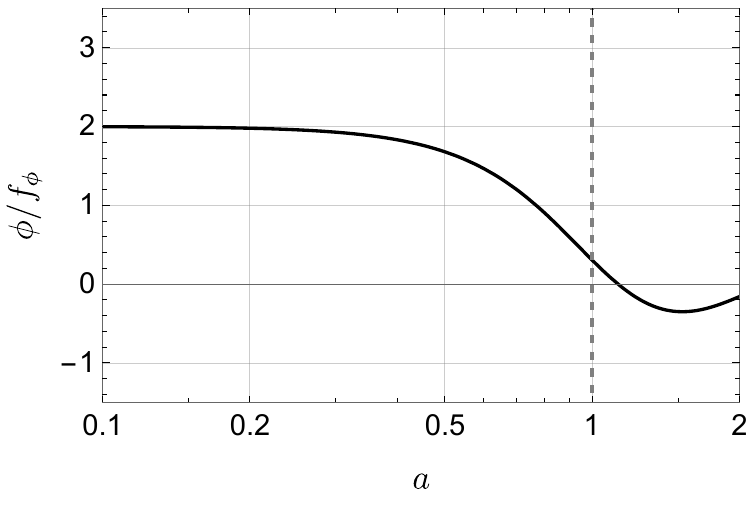}
     \vspace{0.3cm}\\
     \includegraphics[width=1.\linewidth]{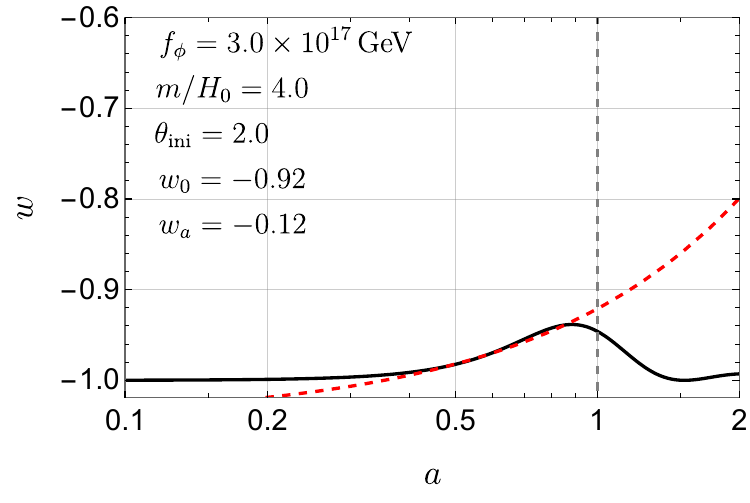}
	\caption{
        Time evolution of the axion field value 
        (top panel) and its equation of state (bottom panel) for the parameter set: $f_\phi = 3.0 \times 10^{17} \GeV$, $m = 4.0 H_0$, and $\theta_{\rm ini} = 2.0$. 
        The red dashed line is the CPL parametrization with $w_0=-0.92$ and $w_a=-0.12$ according to Eq.~\eqref{eq:CPL}
    }
	\label{fig:evolution}
\end{figure}

Figure~\ref{fig:evolution} shows an example for the time evolution of the axion field for the parameter set: 
$f_\phi = 3.0 \times 10^{17} \GeV$, $m = 4.0 H_0$, and $\theta_{\rm ini} = 2.0$. 
The vertical dashed line indicates the present time ($a=1$). 
We set the density parameter for the cosmological constant to be
$\Omega_\Lambda = 0.678$. 
The field displacement until today is given by $\Delta \phi/f_\phi \simeq 1.7$, which corresponds to a rotation angle of $\beta \simeq 0.3^\circ$ for $c_\gamma = 2.6$. The bottom panel of Fig.~\ref{fig:evolution} shows the corresponding evolution of the axion equation of state.  
To facilitate comparison with 
{recent data},
we fit the evolution using the conventional Chevallier–Polarski–Linder (CPL) parametrization~\cite{Chevallier:2000qy,Linder:2002et},
\beq
 w(a) = w_0 + w_a (1-a) \ ,
 \label{eq:CPL}
\eeq
within $a \in (0.45,0.625)$ that corresponds to the redshift range for BAO-optimized sample of DES, $z \in (0.6,1.2)$. 
Since we are considering a time-dependent equation of state, this parametrization allows a reasonable comparison with results presented in Ref.~\cite{DES:2025bxy}. 
For the parameter set shown in Fig.~\ref{fig:evolution}, we obtain 
$(w_0,w_a) = (-0.92,-0.12)$, which is consistent with BAO$+$SN$+$BBN$+\theta_*+t_U$ combined data.

\begin{figure}[t]
	\centering	
     \includegraphics[width=1.\linewidth]{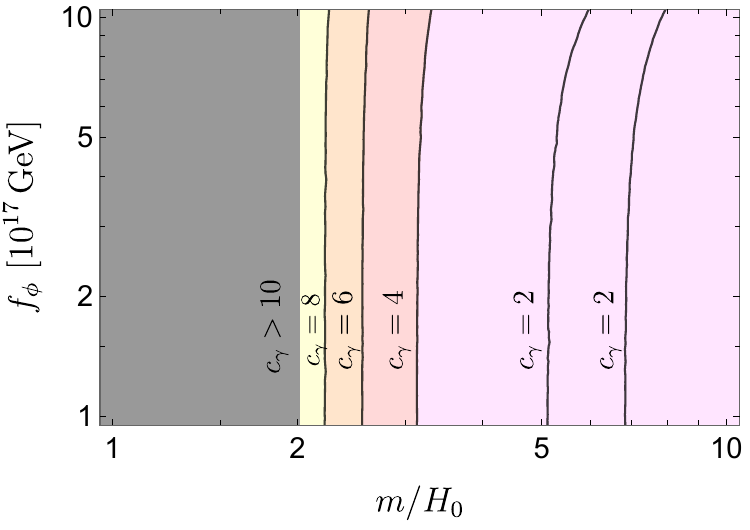}
     \vspace{0.2cm}
     \\
     \includegraphics[width=1.\linewidth]{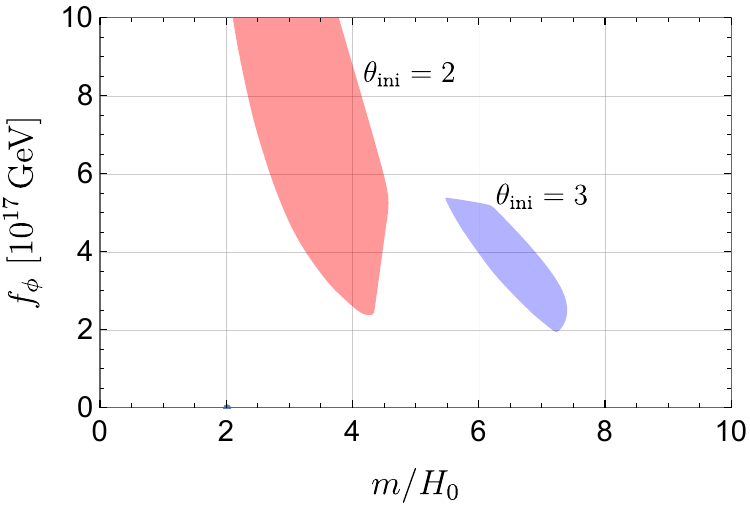}
	\caption{
        {\bf (Upper panel:)} Contour for $c_\gamma$ required to explain $\beta = 0.3^{\circ}$ in $f_\phi\,\text{-}\,m$ plane for $\theta_{\rm ini} = 2$. 
        The cosmological constant is chosen within the range $\Omega_\Lambda \in (0.3,0.7)$ to reproduce the correct present-day relic abundance. 
        The gray-shaded region is disfavored since it corresponds to $c_\gamma > 10$. 
        The value of $c_\gamma$ reaches its minimum around $m / H_0 \sim 6$ and increases for both larger and smaller values of $m/H_0$. 
        {\bf (Lower panel:)} Predicted values in the $m\,\text{-}\,f_\phi$ plane for $\theta_{\rm ini} =2$ (red) and $3$ (blue), consistent with dark energy within the $1\sigma$ region favored by BAO$+$SN$+$BBN$+\theta_*+t_U$ combined data~\cite{DES:2025bxy}. 
    }
	\label{fig:ICB}
\end{figure}

We then scan the parameter set by changing $f_\phi$ and $m$. We take $\theta_{\rm ini} = 2.0$ as an example. 
For each parameter set, we adjust $\Omega_\Lambda$ such that the total energy density from the cosmological constant and the axion lies close to $1-\Omega_{\rm mat} \simeq 0.7$. 
The axion-photon coupling $c_\gamma$ is determined so that the induced cosmic birefringence angle is fixed at $\beta = 0.3^\circ$.

The upper panel of Fig.~\ref{fig:ICB} shows the result for $c_\gamma$ in $f_\phi\,\text{-}\,m$ plane for $\theta_{\rm ini} = 2$.
Here, we choose the range $\Omega_\Lambda \in (0.3, 0.7)$ in order to reproduce the correct present-day relic abundance for $f_\phi \in (10^{17}, 10^{18}) \GeV$, as can be inferred from \FIG{fig:DEratio}.
The gray shaded region is not favored because $c_\gamma > 12$ for such a small axion mass range. 
We find that $c_\gamma$ is as small as $\mathcal{O}(1)$ for $m \gtrsim 2H_0$.

\begin{figure}[t]
	\centering	\includegraphics[width=1.\linewidth]{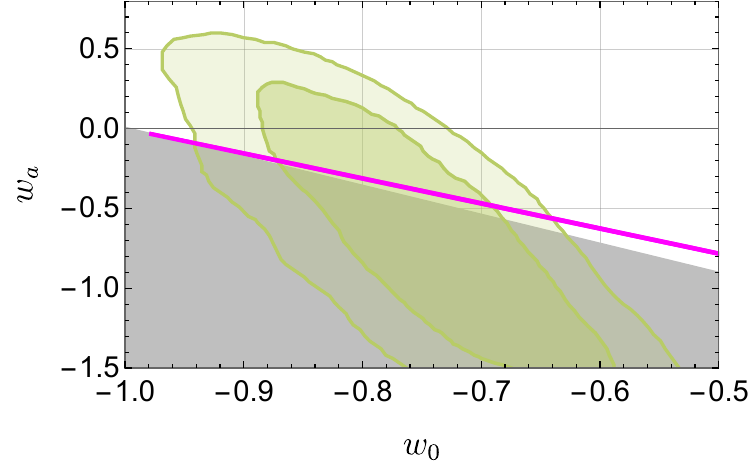}
	\caption{
        Prediction for $w_a\,\text{-}\,w_0$ under CPL parametrization~\eqref{eq:CPL} during $0.45<a<0.625$ (pink line).
        The green (light green) contour represents $1\sigma$ ($3\sigma$) region favored by BAO$+$SN$+$BBN$+\theta_*+t_U$ combined data~\cite{DES:2025bxy}. 
    }
	\label{fig:scatter1}
\end{figure}

Figure~\ref{fig:scatter1} shows our prediction for $w_a$ vs. $w_0$
during $0.45<a<0.625$.
The contour regions represent the parameter space favored by BAO$+$SN$+$BBN$+\theta_*+t_U$ combined data~\cite{DES:2025bxy}. 
The gray shaded region, which represents $w_0 + w_a(1-a_*)$ with $a_* = 0.45$ is smaller than $-1$, is theoretically unphysical because of ``phantom" behavior in the range for BAO-optimized sample of DES, $z \in (0.6,1.2)$.

The lower panel of Fig.~\ref{fig:ICB} shows the parameter regions in which the equation of state is consistent with DE within the $1\sigma$ region favored by the combined BAO$+$SN$+$BBN$+\theta_*+t_U$ data. We consider two cases: $\theta_{\rm ini} = 2$ (red region) and $3$ (blue region).
In the latter case, the anharmonic effect near the top of the sine-Gordon potential (around $\theta_{\rm ini} \sim \pi$) delays the onset of oscillations, thereby restoring consistency with observations even for relatively large values of $m$.
In both scenarios, the axion decay constant $f_\phi$ can be as small as $2 \times 10^{17}$~GeV.

We note that the choice of the axion mass $m$ plays a crucial role in this model.  
For $m \lesssim 2 H_0$, the axion stays nearly frozen and fails to generate notable birefringence or the observed dark energy. 
Conversely, if $m \gtrsim 7 H_0$, the axion begins to roll too early, resulting in a negligible deviation from $w = -1$ at the present epoch. 
Taking $H_0 \simeq 1.5\times 10^{-33}$~eV, we predict 
\begin{equation}
    3.0\times 10^{-33}\,{\rm eV} \lesssim m \lesssim 1.1\times 10^{-32}\,{\rm eV}\;,
\end{equation}
with $f_\phi \sim \mathcal{O}(10^{17})$~GeV and $c_\gamma \sim \mathcal O(1)$.

\section{Discussion and Conclusion
\label{sec:Discussion}}

We have proposed an axion model that simultaneously accounts for both ICB and evolving DE.
In contrast to conventional axion dark energy and quintessence models, our framework includes a constant vacuum energy, which contributes to modifying the equation of state at late times.
Since the axion energy density is subdominant, the resulting equation of state exhibits a mild deviation from $-1$, in agreement with the DESI observations.

Compared to models with only a constant axion mass, our scenario accommodates cosmic birefringence within a broader parameter space using $\mathcal{O}(1)$ parameters, particularly when the axion-photon coupling involves an $\mathcal{O}(1)$ anomaly coefficient. 
Moreover, since the axion field does not need to account for the entire DE density, its decay constant (or equivalently, its oscillation amplitude) can be significantly smaller than the Planck scale.
This feature is especially favorable for a string axion.

Our model is a minimal extension of the $\Lambda$CDM framework. A skeptical reader might argue that an axion quintessence model is simpler, as it does not explicitly introduce a constant vacuum energy.
However, it is important to recognize that a cosmological constant is generically present in any realistic particle physics model.
In quintessence models, one must assume that this constant vacuum energy is vanishingly small, which constitutes an implicit fine-tuning.
In this respect, our model is no more complex than the axion quintessence model. Moreover, it offers advantages such as allowing for a smaller axion decay constant and an axion-photon coupling of order unity.

We fit our results using the CPL parametrization for the evolving equation of state to facilitate comparison with existing constraints from the latest data. However, the fitting function shows significant deviation at earlier epochs, as our model does not evolve into a ``phantom" phase. A more detailed analysis comparing our predictions with observational data is warranted for a more quantitative study.

\section*{Acknowledgments}
YN is supported by Natural Science Foundation of Shanghai.
MY is supported by JSPS KAKENHI Grant Numbers 20H05851 and 23K13092.
The work of Y. -C. Qiu is supported by the K. C.
Wong Educational Foundation.

\bibliography{reference}

\end{document}